\documentstyle[aps,twocolumn,prb,epsf]{revtex}
\begin{document}
\title{\hspace{8cm} {\em to appear in JETP}\\
\vskip 0.25cm
Probing the field-induced variation of the chemical potential
in $Bi_2Sr_2CaCu_2O_y$ via the magneto-thermopower measurements}
\author{S.A. Sergeenkov}
\address{SUPRAS, Institute of Physics, University of Liege,
B-4000, Liege, Belgium;\\
Bogoliubov Laboratory of Theoretical Physics,
Joint Institute for Nuclear Research,\\ 141980 Dubna, Moscow Region, Russia}
\author{M. Ausloos}
\address{SUPRAS, Institute of Physics, University of Liege,
B-4000, Liege, Belgium}
\date{\today}
\maketitle
\begin{abstract}
Approximating the shape of the measured in textured $Bi_2Sr_2CaCu_2O_y$
magneto-thermopower (TEP) $\Delta S(T,H)$ by asymmetric linear
triangle of the form $\Delta S(T,H)\simeq S_p(H)\pm B^{\pm}(H)(T_c-T)$ with
positive $B ^{-}(H)$ and $B ^{+}(H)$ defined below and above $T_c$, we
observe that $B ^{+}(H)\simeq 2B ^{-}(H)$. In order to account for this asymmetry,
we explicitly introduce the field-dependent chemical potential of holes
$\mu (H)$ into
the Ginzburg-Landau theory and calculate both an average $\Delta S_{av}(T,H)$
and fluctuation $\Delta S_{fl}(T,H)$ contributions to the total magneto-TEP
$\Delta S(T,H)$. As a result, we find a rather simple relationship between
the field-induced variation of the chemical potential
in this material and the above-mentioned magneto-TEP data around $T_c$,
viz. $\Delta \mu (H)\propto S_p(H)$.
\end{abstract}

\narrowtext

As is well-known,~\cite{1,2} the variation of the chemical potential $\mu$ of
free carriers in an applied magnetic field $H$ provides a direct information
about the magnetization structure inside a superconducting sample. Namely,
the field-induced change of the chemical potential in superconducting state
reads~\cite{3} $\Delta \mu (H)\equiv \mu (H)-\mu (0)=-M(H)H/n$,
where $M(H)$ is the field-induced magnetization, and $n$ the carrier number
density.
At the same time, due to the existence of the so-called compensation
effect,~\cite{4} it is rather difficult to observe field-induced
modulations of $\mu$ in bulk samples since in equilibrium any
field-induced variations of $\mu$ will be
completely canceled by similar variations caused by the magnetostrictive
changes of the volume. However, this compensation does not occur in thin
films~\cite{1,2} and oriented powders.~\cite{5} And thus we can expect to
see some tangible changes of $\mu (H)$ in layered (anisotropic) structures
as well. On the other hand, in view of its carrier sensitive
nature, thermopower (TEP) measurements seem to be the most adequate tool for
probing the field-induced changes of the chemical potentials.
Indeed, TEP results have already proved to be useful for providing reasonable
estimates for such important physical parameters
as the Fermi energy, Debye temperature, interlayer spacing etc.~\cite{6,7}
Studying the observable magneto-TEP
$\Delta S(T,H)=S(T,H)-S(T,0)$ also provides important insights into different
aspects of the material in the mixed state~\cite{7,8,9}
(when $H_{c1}\ll H\ll H_{c2}$).
When experimental results are presented in the form of the above-defined
$\Delta S(T,H)$ one observes that its temperature dependence has a
$\Lambda$-like shape asymmetric around $T_c$ where it reaches its magnetic
field-dependent peak value $S_p(H)\equiv \Delta S(T_c,H)$. Then, for small
fields, approximating the shape of $\Delta S(T,H)$ by the asymmetric linear
triangle of the form~\cite{8}
\begin{equation}
\Delta S(T,H)\simeq S_p(H)\pm B^{\pm}(H)(T_c-T),
\end{equation}
with positive slopes $B ^{-}(H)$ and $B ^{+}(H)$ defined for $T<T_c$ and
$T>T_c$, respectively, one finds (see Fig.1) that $B^{+}(H)\simeq 2B^{-}(H)$
in the vicinity of $T_c$.

In the present paper, using the Ginzburg-Landau theory and utilizing some
typical magneto-TEP data~\cite{7,8} on textured $Bi_2Sr_2CaCu_2O_y$, we
discuss the mixed-state behavior of the magneto-TEP
(and in particular the origin of the asymmetry given by Eq.(1)) via the
corresponding behavior of the chemical potential in applied magnetic field.

It is well-known~\cite{7,8,9} that for external fields $H$ such that
$H_{c1}\ll H\ll H_{c2}$ and for the Ginzburg-Landau parameter $\kappa \gg 1$,
the magneto-TEP $\Delta S(T,H)$ is proportional to the strength of the
external field. To describe the observed behavior of the magneto-TEP
both below and above $T_c$, we can roughly present it in a two-term
contribution form~\cite{7}
\begin{equation}
\Delta S(T,H)=\Delta S_{av}(T,H)+\Delta S_{fl}(T,H),
\end{equation}
where the average term $\Delta S_{av}(T,H)$ is assumed to be non-zero only
below $T_c$ (since in the normal state the TEP of high-$T_c$ superconductors
(HTS) is found to be
very small~\cite{8,9}) while the fluctuation term $\Delta S_{fl}(T,H)$
should contribute to the observable $\Delta S(T,H)$ for $T\simeq T_c$.
In what follows, we shall discuss these two contributions separately within
a mean-field theory approximation.

\paragraph{Mean value of the magneto-TEP: $\Delta S_{av}(T,H)$.}
Assuming that the net result of the magnetic field is to modify the
chemical potential (Fermi energy) $\mu$ of quasiparticles, we can write the
the generalized GL free energy functional ${\cal G}$ of a superconducting
sample in the mixed state as
\begin{equation}
{\cal G} [\psi ]=a(T)|\psi |^2+\frac{\beta}{2}|\psi |^4-\mu |\psi |^2.
\end{equation}
Here $\psi =|\psi |e^{i\phi}$ is the superconducting order parameter,
$\mu (H)$ stands for the field-dependent in-plane chemical potential
of quasiparticles; $a(T,H)=\alpha (H)(T-T_c)$ and the GL parameters
$\alpha (H)$ and $\beta (H)$ are related to the critical temperature $T_c$,
zero-temperature BCS gap $\Delta _0=1.76k_BT_c$, the out-of-plane chemical
potential (Fermi energy) $\mu _c(H)$, and the total particle number density
$n$ as $\alpha (H)=\beta (H)n/T_c=2\Delta _0k_B/\mu _{c}(H)$. In fact, in
layered superconductors, $\mu =\mu _c/\gamma ^2\simeq m^{*}_{ab}
(J_cd/2\hbar )^2$, where $d$ and $J_c$ are the interlayer distance and
coupling energy within the Lawrence-Doniach model, 
and $\gamma =\sqrt{m_c^{*}/m_{ab}^{*}}$ is the mass anisotropy ratio. The
magnetic field is applied normally to the $ab$-plane where
the strongest magneto-TEP effects are expected.~\cite{9} In what follows,
we ignore the field dependence of the critical temperature since for all
fields under discussion $T_c(H)=T_c(0)(1-H/H_{c2})\simeq T_c(0)\equiv T_c$.

As usual, the equilibrium state of such a system is determined from the
minimum energy condition $\partial {\cal G}/\partial |\psi |=0$ which
yields for $T<T_c$
\begin{equation}
|\psi _0|^2=\frac{\alpha (H)(T_c-T)+\mu (H)}{\beta (H)}
\end{equation}
Substituting $|\psi _0|^2$ into Eq.(3) we obtain for the average free
energy density
\begin{equation}
\Omega (T,H)\equiv {\cal G} [\psi _0]=-
\frac{[\alpha (H)(T_c-T)+\mu (H)]^2}{2\beta (H)}
\end{equation}
In turn, the magneto-TEP $\Delta S(T,H)$ can be
related to the
corresponding difference of transport entropies~\cite{7,8}
$\Delta \sigma \equiv
\partial \Delta \Omega /\partial T$ as $\Delta S(T,H)=\Delta \sigma (T,H)/en$,
where $e$ is the charge of the quasiparticles.
Finally the mean value of the mixed-state magneto-TEP reads (below $T_c$)
\begin{equation}
\Delta S_{av}(T,H)=S_{p,av}(H)-B _{av}(H)(T_c-T),
\end{equation}
with
\begin{equation}
S_{p,av}(H)=\frac{\Delta \mu (H)}{eT_c},
\end{equation}
and
\begin{equation}
B _{av}(H)=\frac{8\Delta _0k_B\Delta \mu (H)}{eT_c\gamma ^2\mu ^2(0)}.
\end{equation}
Before we proceed to compare the above theoretical findings with the
available experimental
data, we first have to estimate the corresponding fluctuation
contributions to the observable magneto-TEP, both above and below $T_c$.

\paragraph{Mean-field Gaussian fluctuations of the magneto-TEP:
$\Delta S_{fl}(T,H)$.}
The influence of superconducting fluctuations on transport properties of HTS
(including TEP and electrical conductivity) has been extensively
studied for the past few years (see, e.g.,~\cite{10,11,12,13,14} and further
references therein). In particular, it was found that the
fluctuation-induced behavior may extend to temperatures more than
$10K$ higher than the respective $T_c$.
Let us consider now the region near $T_c$ and discuss the Gaussian
fluctuations of the mixed-state magneto-TEP $\Delta S_{fl}(T,H)$.
Recall that according to the theory of Gaussian fluctuations,~\cite{15}
the fluctuations of any observable, which is conjugated to the order
parameter $\psi$ (such as heat
capacity, susceptibility, etc) can be presented in terms of the statistical
average of the square of the fluctuation amplitude $<(\delta \psi )^2>$ with
$\delta \psi =\psi -\psi _0$. Then the TEP above $(+)$ and
below $(-)$ $T_c$ have the form of
\begin{equation}
S_{fl}^{\pm}(T,H)=A<(\delta \psi )^2>_{\pm}
=\frac{A}{Z}\int d|\psi |(\delta \psi )^2 e^{-\Sigma [\psi ]},
\end{equation}
where
$Z=\int d|\psi |e^{-\Sigma [\psi ]}$ is the partition function with
$\Sigma [\psi ]\equiv ({\cal G} [\psi ]-{\cal G} [\psi _0])/k_BT$, and $A$
is a coefficient to be defined below.
Expanding the free energy density functional ${\cal G} [\psi ]$
\begin{equation}
{\cal G} [\psi ]\approx {\cal G} [\psi _0]+
\frac{1}{2}\left[ \frac{\partial ^2{\cal G}}
{\partial \psi ^2}\right ]_{|\psi |=|\psi _0|}\!(\delta \psi )^2,
\end{equation}
around the mean value of the order parameter $\psi _0$, which is defined as a
stable solution of equation $\partial {\cal G}/\partial |\psi |=0$ we can
explicitly calculate the Gaussian integrals.
Due to the fact that $|\psi _0|^2$ is given by Eq.(4) below $T_c$ and
vanishes at $T\ge T_c$, we obtain finally
\begin{equation}
S_{fl}^{-}(T,H)=\frac{Ak_BT_c}{4\alpha (H)(T_c-T)+4\mu (H)},
\qquad T\le T_c
\end{equation}
and
\begin{equation}
S_{fl}^{+}(T,H)=\frac{Ak_BT_c}{2\alpha (H)(T-T_c)-2\mu (H)},
\qquad T\ge T_c
\end{equation}
As we shall see below, for the experimental range of parameters under
discussion, $\mu (H)/\alpha (H)\gg |T_c-T|$. Hence, with a good
accuracy we can linearize Eqs.(11) and (12) and obtain for the fluctuation
contribution to the magneto-TEP
\begin{equation}
\Delta S_{fl}^{\pm}(T,H)\simeq S_{p,fl}^{\pm}(H)\pm B _{fl}^{\pm}(H)(T_c-T),
\end{equation}
where
\begin{equation}
S_{p,fl}^{-}(H)=-\frac{Ak_BT_c\Delta \mu (H)}{4\mu ^2(0)},
\quad S_{p,fl}^{+}(H)=-2S_{p,fl}^{-}(H),
\end{equation}
and
\begin{equation}
B _{fl}^{-}(H)=-\frac{3Ak_B^2T_c\Delta _0\Delta \mu (H)}{\gamma ^2\mu ^4(0)},
\quad B _{fl}^{+}(H)=-2B _{fl}^{-}(H).
\end{equation}
Furthermore, it is reasonable to assume that $S_p^{-}=S_p^{+}\equiv 
S_p$, where $S_{p}^{-}=S_{p,av}+S_{p,fl}^{-}$ and $S_{p}^{+}=S_{p,fl}^{+}$. 
Then the above equations
bring about the following explicit expression for the constant
parameter $A$, namely $A=4\mu ^2(0)/3ek_BT_c^2$. This in turn leads
to the following expressions for the fluctuation contribution to
peaks and slopes through their average counterparts (see Eqs.(7) and (8)):
$S_{p,fl}^{+}(H)=(2/3)S_{p,av}(H)$, $S_{p,fl}^{-}(H)=-(1/3)S_{p,av}(H)$,
$B _{fl}^{-}(H)=-(1/2)B_{av}(H)$, and $B _{fl}^{+}(H)=B_{av}(H)$.
Finally, the total contribution to the observable magneto-TEP reads
(Cf. Eq.(1))
\begin{equation}
\Delta S(T,H)=S_p(H)\pm B^{\pm}(H)(T_c-T),
\end{equation}
where
\begin{equation}
S_{p}(H)=\frac{2\Delta \mu (H)}{3eT_c}, \quad B^{+}(H)\equiv B _{fl}^{+}(H)
=2B^{-}(H),
\end{equation}
and
\begin{equation}
B^{-}(H)\equiv B_{av}(H)+B^{-}_{fl}(H)=\frac{4\Delta _0k_B\Delta \mu (H)}
{eT_c\gamma ^2\mu ^2(0)}.
\end{equation}
Let us compare now the obtained theoretical expressions with the typical
experimental data~\cite{8} on textured $Bi_2Sr_2CaCu_2O_y$ for the slopes
$B ^{\pm}(H)$ and the
peak $S_p(H)$ values for $H=0.12T$ (see Fig.1): $S_p=0.16\pm 0.01 \mu V/K$,
$B ^{-}=0.012\pm 0.001 \mu V/K^2$, and $B^{+}=0.027\pm 0.003 \mu V/K^2$.
First we notice that the calculated
slopes $B ^{+}(H)$ above $T_c$ are twice their counterparts below $T_c$,
i.e., $B ^{+}(H)=2B ^{-}(H)$ in a good agreement with the observations.
Using $\gamma \simeq 55$ and $d=1.2nm$ for the
anisotropy ratio and interlayer distance in this material,~\cite{9,13,16} we
obtain reasonable estimates of the field-induced changes of the in-plane
chemical potential (Fermi energy) $\Delta \mu (H)$ (along with its
zero-field value $\mu (0)$) and the interlayer
coupling energy $J_c$. Namely, $\mu (0)\simeq 1.6meV$,
$\Delta \mu (H)\simeq 0.02meV$, and $J_c\simeq 4meV$. Furthermore,
relating the field-induced variation of the in-plane chemical potential
to the change of the corresponding magnetization $M(H)$, viz.
\begin{equation}
\Delta \mu (H)=-\frac{M(H)H}{n_h},
\end{equation}
where $M(H)$ for $H_{c1}\ll H\ll H_{c2}$ has a form~\cite{3} (recall that the
lower critical field for this material is
$H_{c1}=(\phi _0/4\pi \lambda _{ab}^2)\ln {\kappa}\simeq 40G$ with
$\lambda _{ab}\simeq 250nm$, $\xi _{ab}\simeq 1nm$, and $\kappa \simeq 250$)
\begin{equation}
\mu _0M(H)=\frac{2\phi _0}{\sqrt{3}\lambda _{ab}^2}
\left \{\ln \left [\frac{3\phi _0}{4\pi \lambda _{ab}^2(H-H_{c1})}\right ]
\right \}^{-2}-H,
\end{equation}
we obtain $n_h\simeq 2.5\times 10^{27}m^{-3}$ for the hole number density
in this material, in reasonable agreement with the other estimates of this
parameter.~\cite{17} Fig.2 shows $\Delta \mu (H)$ calculated according to
Eq.(19) with
the experimental data points deduced (via Eq.(17)) from the magneto-TEP
measurements on the same sample.~\cite{7} As is seen,
the data are in a good agreement with the model predictions.
And finally, using the above parameters (along with the critical temperature),
we find that $\mu (H)/\alpha (H)\simeq 100K$ which justifies the use
of the linearized Eq.(13) since, as is seen in Fig.1, the observed
magneto-TEP practically vanishes for $|T_c-T|\ge 15K$.

In conclusion, to probe the variation of chemical potential $\Delta \mu (H)$
of quasiparticles in anisotropic materials
under an applied magnetic field, we calculated the mixed-state
magneto-thermopower $\Delta S(T,H)$ in the presence of field-modulated
charge effects near $T_c$. Using the available magneto-TEP experimental data
on textured $Bi_2Sr_2CaCu_2O_y$, field-induced behavior of in-plane
$\Delta \mu (H)$
was obtained along with reasonable estimates for its zero-field value
(Fermi energy) $\mu (0)$, interlayer coupling energy $J_c$, and the hole
number density $n_h$ in this material.

We thank A. Varlamov for very useful discussions on the subject.
Part of this work has been financially supported by the Action de Recherche
Concert\'ees (ARC) 94-99/174. S.A.S. acknowledges
the financial support from FNRS.

\begin{figure}[htb]
\epsfxsize=8cm
\centerline{\epsffile{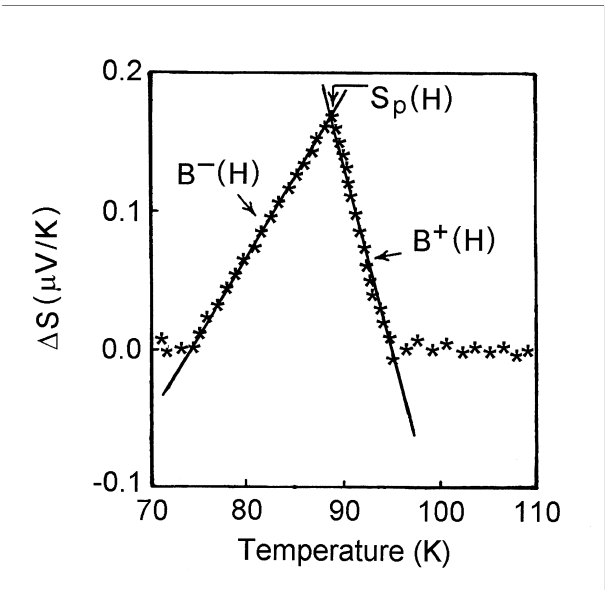} }
\caption{A typical pattern of the observed~\cite{8} magneto-TEP 
$\Delta S(T,H)$ of textured $Bi_2Sr_2CaCu_2O_y$ at $H=0.12T$. The best
fit to the data points according to Eq.(1) yields $S_p(H)=0.16\pm 0.01 \mu V/K$,
$B^{-}(H)=0.012\pm 0.001 \mu V/K^2$, and $B^{+}(H)=0.027\pm 0.003 \mu V/K^2$ 
for the peak and slopes.}
\end{figure}

\begin{figure}[htb]
\epsfxsize=8cm
\centerline{\epsffile{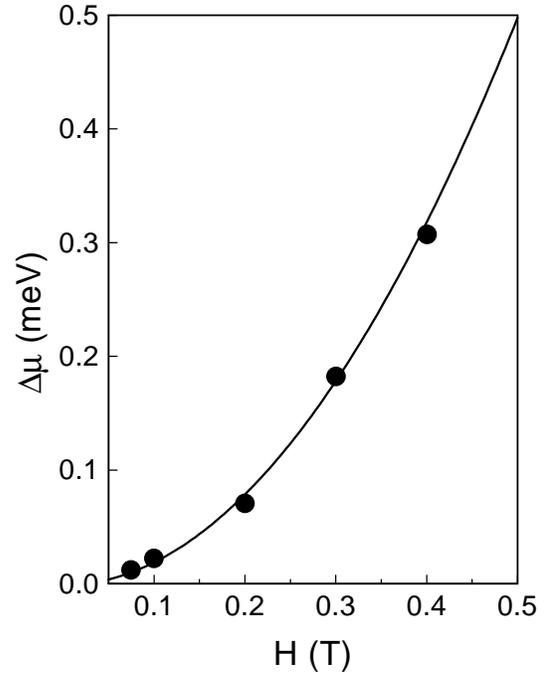} }
\caption{The change of the chemical potential $\Delta \mu (H)$ in applied
magnetic field calculated according to Eq.(19). The experimental points are
deduced from the magneto-TEP data~\cite{7} on $Bi_2Sr_2CaCu_2O_y$ and related
to $\Delta \mu (H)$ via Eq.(17).}
\end{figure}


\begin{thebibliography}{99}
\bibitem{1} V.I. Nizhankovskii, R.N. Sheftal', and S.G. Zybtsev,
JETP Lett. {\bf 55}, 238 (1992).
\bibitem{2} V.I. Nizhankovskii and S.G. Zybtsev, Phys. Rev. B {\bf 50},
1111 (1994).
\bibitem{3} A.A. Abrikosov, {\it Fundamentals of the Theory of Metals},
Elsevier, Amsterdam, 1988.
\bibitem{4} N.E. Alekseevskii and V.I. Nizhankovskii, Sov. Phys. JETP
{\bf 61}, 1051 (1985).
\bibitem{5} C.M. Fowler, B.L. Freeman, W.L. Hults, J.C. King, F.M. Muller,
and J.L. Smith, Phys. Rev. Lett. {\bf 68}, 534 (1992).
\bibitem{6} V. Gridin, S. Sergeenkov, R. Doyle, P. de Villiers, and
M. Ausloos, Phys. Rev. B {\bf 47}, 14594 (1993).
\bibitem{7} S. Sergeenkov, M. Ausloos, H. Bougrine, R. Cloots, and V. Gridin,
Phys. Rev. B {\bf 48}, 16680 (1993).
\bibitem{8} V. Gridin, P. Pernambuco-Wise, C.G. Trendall, W.R. Datars, and
J.D. Garrett, Phys. Rev. B {\bf 40}, 8814 (1989).
\bibitem{9} N.V. Zavaritskii, A.V. Samoilov, and A.A. Yurgens,
JETP Lett. {\bf 55}, 127 (1992).
\bibitem{10} L. Reggiani, R. Vaglio, and A.A. Varlamov, Phys. Rev. B {\bf 44},
9541 (1991).
\bibitem{11} A.A. Varlamov, D.V. Livanov, and F. Federici,
JETP Lett. {\bf 65}, 196 (1997).
\bibitem{12} M. Houssa, H. Bougrine, S. Stassen, R. Cloots, and M. Ausloos,
Phys. Rev. B {\bf 54}, R6885 (1996).
\bibitem{13} M. Houssa, M. Ausloos, R. Cloots, and H. Bougrine,
Phys. Rev. B {\bf 56}, 802 (1997).
\bibitem{14} A. A. Varlamov and M. Ausloos, in {\it Fluctuation Phenomena in
High Temperature Superconductors}, edited by M.Ausloos and A. A. Varlamov,
vol. 32 in the NATO ASI Partnership Sub-Series (Kluwer, Dordrecht, 1997), p. 3.
\bibitem{15} H.E. Stanley, {\it Introduction to Phase Transitions and
Critical Phenomena}, Clarendon Press, Oxford, 1968.
\bibitem{16} W.C. Lee, R.A. Klemm, and D.C. Johnston, Phys. Rev. Lett.
{\bf 63}, 1012 (1989).
\bibitem{17} Xin-Fen Chen, G.X. Tessema, and M.J. Skove, Phys. Rev. B
{\bf 48}, 13141 (1993).
\end{thebibliography}
\end{document}